\documentclass[floatfix,aps,prl,twocolumn,superscriptaddress]{revtex4}
\usepackage{graphicx}
\usepackage[]{epsfig}
\usepackage{times,amsmath,amssymb}

\usepackage{color}

\begin{document}

\title{Tunneling induced spin dynamics in a quantum dot-lead hybrid system}

\author{Tomohiro Otsuka}
\email[]{tomohiro.otsuka@riken.jp}
\affiliation{Center for Emergent Matter Science, RIKEN, 2-1 Hirosawa, Wako, Saitama 351-0198, Japan}
\affiliation{Department of Applied Physics, University of Tokyo, Bunkyo, Tokyo 113-8656, Japan}

\author{Takashi Nakajima}
\affiliation{Center for Emergent Matter Science, RIKEN, 2-1 Hirosawa, Wako, Saitama 351-0198, Japan}
\affiliation{Department of Applied Physics, University of Tokyo, Bunkyo, Tokyo 113-8656, Japan}

\author{Matthieu R. Delbecq}
\affiliation{Center for Emergent Matter Science, RIKEN, 2-1 Hirosawa, Wako, Saitama 351-0198, Japan}

\author{Shinichi Amaha}
\affiliation{Center for Emergent Matter Science, RIKEN, 2-1 Hirosawa, Wako, Saitama 351-0198, Japan}

\author{Jun Yoneda}
\affiliation{Center for Emergent Matter Science, RIKEN, 2-1 Hirosawa, Wako, Saitama 351-0198, Japan}
\affiliation{Department of Applied Physics, University of Tokyo, Bunkyo, Tokyo 113-8656, Japan}

\author{Kenta Takeda}
\affiliation{Center for Emergent Matter Science, RIKEN, 2-1 Hirosawa, Wako, Saitama 351-0198, Japan}

\author{Giles Allison}
\affiliation{Center for Emergent Matter Science, RIKEN, 2-1 Hirosawa, Wako, Saitama 351-0198, Japan}

\author{Peter Stano}
\affiliation{Center for Emergent Matter Science, RIKEN, 2-1 Hirosawa, Wako, Saitama 351-0198, Japan}
\affiliation{Institute of Physics, Slovak Academy of Sciences, 845 11 Bratislava, Slovakia}

\author{Akito Noiri}
\affiliation{Center for Emergent Matter Science, RIKEN, 2-1 Hirosawa, Wako, Saitama 351-0198, Japan}
\affiliation{Department of Applied Physics, University of Tokyo, Bunkyo, Tokyo 113-8656, Japan}

\author{Takumi Ito}
\affiliation{Center for Emergent Matter Science, RIKEN, 2-1 Hirosawa, Wako, Saitama 351-0198, Japan}
\affiliation{Department of Applied Physics, University of Tokyo, Bunkyo, Tokyo 113-8656, Japan}

\author{Daniel Loss}
\affiliation{Center for Emergent Matter Science, RIKEN, 2-1 Hirosawa, Wako, Saitama 351-0198, Japan}
\affiliation{Department of Physics, University of Basel, Klingelbergstrasse 82, 4056 Basel, Switzerland}

\author{Arne Ludwig}
\affiliation{Angewandte Festk\"orperphysik, Ruhr-Universit\"at Bochum, D-44780 Bochum, Germany}

\author{Andreas D. Wieck}
\affiliation{Angewandte Festk\"orperphysik, Ruhr-Universit\"at Bochum, D-44780 Bochum, Germany}

\author{Seigo Tarucha}%
\affiliation{Center for Emergent Matter Science, RIKEN, 2-1 Hirosawa, Wako, Saitama 351-0198, Japan}
\affiliation{Department of Applied Physics, University of Tokyo, Bunkyo, Tokyo 113-8656, Japan}
\affiliation{Quantum-Phase Electronics Center, University of Tokyo, Bunkyo, Tokyo 113-8656, Japan}
\affiliation{Institute for Nano Quantum Information Electronics, University of Tokyo, 4-6-1 Komaba, Meguro, Tokyo 153-8505, Japan}

\date{\today}
\begin{abstract}
Semiconductor quantum dots (QDs) offer a platform to explore the physics of quantum electronics including spins.
Electron spins in QDs are considered good candidates for quantum bits~\cite{1998LossPRA} in quantum information processing~\cite{2000NielsenBk, 2010LaddNat},
and spin control and readout have been established down to a single electron level~\cite{2007HansonRMP}. 
We use these techniques to explore spin dynamics in a hybrid system, namely a QD coupled to a two dimensional electronic reservoir. 
The proximity of the lead results in relaxation dynamics of both charge and spin, the mechanism of which is revealed by comparing the charge and spin signal. For example, higher order charge tunneling events can be monitored by observing the spin.
We expect these results to stimulate further exploration of spin dynamics in QD-lead hybrid systems and expand the possibilities for controlled spin manipulations.
\end{abstract}

\maketitle



Electron spins in semiconductor QDs have relatively long coherence times~\cite{2011BluhmNatPhys, 2014ShulmanNatCom, 2014KawakamiNatNano, 2014VeldhorstNatNano}, while the solid state structures have potential scalability by utilizing current extensive semiconductor fabrication techniques.
They are also considered good candidates for quantum bits~\cite{1998LossPRA} in quantum information processing~\cite{2000NielsenBk, 2010LaddNat}.
The spin states can be initialized using a large exchange energy of a single QD~\cite{2005PettaSci}, and the related Pauli spin blockade in a double QD (DQD)~\cite{2002OnoSci}.
The demonstrated ways of manipulation include the spin-spin exchange interaction between neighboring QDs~\cite{2005PettaSci}, and electron spin resonances induced by micro coils~\cite{2006KoppensNat}, nuclear spins~\cite{2007LairdPRL}, spin orbit interaction~\cite{2007NowackSci}, or micro magnets~\cite{2006TokuraPRL, 2008LadriereNatPhys, 2014YonedaPRL}.
Finally, the spin can be read out by Pauli spin blockade, or a tunneling sensitive to the Zeeman energy~\cite{2004ElzermanNature}.

In experiments aimed at the minimization of the dissipation, the QDs have been typically isolated from their environment, including the leads, as much as possible. However, it is worth to explore physics in hybrid systems, where the dot-environment coupling is stronger, since this coupling can be tuned straightforwardly and precisely tuned by gate voltages.
In addition, the electronic reservoirs can be tailored themselves, by applying bias voltages or using specific states such as ferromagnets~\cite{2007HamayaAPL}, superconductors~\cite{2010DeaconPRL}, quantum Hall states~\cite{2010AltimirasNatPhys}, and others.
This variability gives rise to attractive physics like Fano interference~\cite{1961FanoPR, 2002KobayashiPRL}, Kondo states~\cite{1998GordonNat, 2000vanderWielSci}, or general physics of open and nonequilibrium systems, and possibly lead to new ways of spin manipulations utilizing interactions induced by the environment~\cite{2006AndersPRB}. 

In this work, we explore spin dynamics in a QD-lead hybrid system utilizing the spin manipulation and readout techniques developed in previous spin qubit experiments.
Specifically, we monitor changes of the spin and charge states induced by coupling of the QD to an electronic reservoir.
With the QD being close to a charge transition, we observe spin and charge relaxation, corresponding to first-order tunneling events.
With the dot in a Coulomb blockade configuration, we observe only the relaxation of spin, corresponding to second-order tunneling events.


\begin{figure}
\begin{center}
  \includegraphics{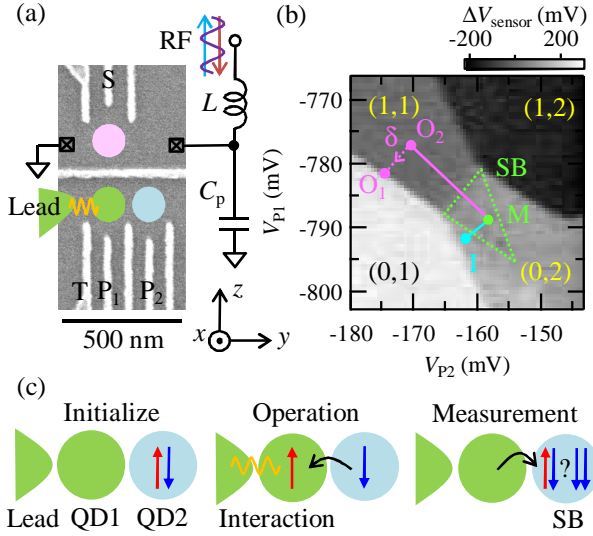}
  \caption{(a) Scanning electron micrograph of the device and the schematic of the measurement setup.
  A DQD is formed at the lower side and the charge states are monitored by the charge sensor QD at the upper side.
  The charge sensor is connected to resonators formed by the inductor $L$ and the stray capacitance $C_{\rm p}$ for the RF reflectometry.
  The external magnetic field of 0.5~T is applied in plane along the $z$ axis.
  (b) $\Delta V_{\rm sensor}$ as a function of $V_{\rm P2}$ and $V_{\rm P1}$.
  Changes of the charge states are observed.
  The number of electrons in each QD is given as $(n_{\rm 1}, n_{\rm 2})$.
  The triangle shows the region of spin blockade.
  The positions corresponding to steps of pulse sequences (O, I, M) are indicated. 
  (c) Schematic of the measurement scheme.
  The spin state is initialized to a (0,2) singlet at I.
  Next, the we move into O in (1,1) where the spin couples to the lead.
  Finally, the spin state is measured using spin blockade at M.
  }
  \label{Device}
\end{center}
\end{figure}

Figure~\ref{Device}(a) shows a scanning electron micrograph of the device.
By applying negative voltages on the gate electrodes, a DQD and a QD charge sensor~\cite{2010BarthelPRB} are formed at the lower and upper sides, respectively.
The left QD in the DQD couples to a lead and the coupling strength is tuned by the voltage $V_{\rm T}$ applied on gate T.
The QD charge sensor is connected to a RF resonator formed by the inductor $L$ and the stray capacitance $C_{\rm p}$ for RF reflectometry~\cite{2010BarthelPRB, 1998SchelkopfSci, 2007ReillyAPL}.
The number of electrons in each QD $(n_{\rm 1}, n_{\rm 2})$ is monitored by the intensity of the reflected RF signal $V_{\rm sensor}$.
A change in the electrostatic environment around the sensing dot changes its conductance, which shifts the tank circuit resonance and modifies $V_{\rm sensor}$ measured at $f_{\rm res}$=297~MHz, the circuit resonance frequency.

Figure~\ref{Device}(b) shows the charge stability diagram of the DQD.
We measured the sensor signal $V_{\rm sensor}$ as a function of the plunger gate voltages of QD$_{2}$ ($V_{\rm P2}$), and QD$_{1}$ ($V_{\rm P1}$).
We observe a change $\Delta V_{\rm sensor}$ each time the DQD charge configuration $(n_{\rm 1}, n_{\rm 2})$ changes. Depicted in Fig.~\ref{Device}(b), the values
$(n_{\rm 1}, n_{\rm 2})$ are assigned by counting the number of charge transition lines from the fully depleted configuration $(n_{\rm 1}, n_{\rm 2})$=(0,0) [the latter not shown on Fig.~\ref{Device}(b)].
Around the charge state transition $(1,1)\leftrightarrow (0,2)$, we observe a suppression of the $(0,2)$ charge signal due to the Pauli spin blockade [in the region indicated by the triangle in Fig.~\ref{Device}(b)]. In this specific measurement of the stability diagram, unlike elsewhere, upon pulsing $(2,0) \rightarrow (1,1)$ we move through the  singlet-triplet $T_+$ anti-crossing very slowly (adiabatically), to induce a sizable triplet component of the $(1,1)$ state even at a zero interaction time. Pulsing quickly back $(1,1) \rightarrow (0,2)$ results in a Pauli blocked signal inside the denoted triangular area. This shows us where we can utilize the Pauli spin blockade to readout the spin state in the following measurements, probing the dot spin and charge tunneling-induced dynamics.

The operation scheme to measure the effect of the lead on the spin is depicted in Fig.~\ref{Device}(c).
We initialize the state to a (0,2) singlet by waiting at the initialization point I denoted in Fig.~\ref{Device}(b).
Next, we move to the operation point O.
In this step, the electron in QD$_{1}$ interacts with the lead and the dot state might be changed by electron tunneling.
The tunneling rate can be modified by tuning $V_{\rm T}$, which changes the tunnel coupling, and the position of O, which changes the dot potential with respect to the Fermi energy of the lead (O$_1$: close to a charge transition, O$_2$: deep in the Coulomb blockade).
At the next step, the spin state is measured using spin blockade by pulsing the dot to the point denoted by M.
If the spin state did not change, we observe the (0,2) singlet again.
If the spin state changed, a polarized triplet component is measured as a blocked $(1,1)\rightarrow (0,2)$ charge transition. From the charge signal, we can therefore deduce the spin state. 


\begin{figure}
\begin{center}
  \includegraphics{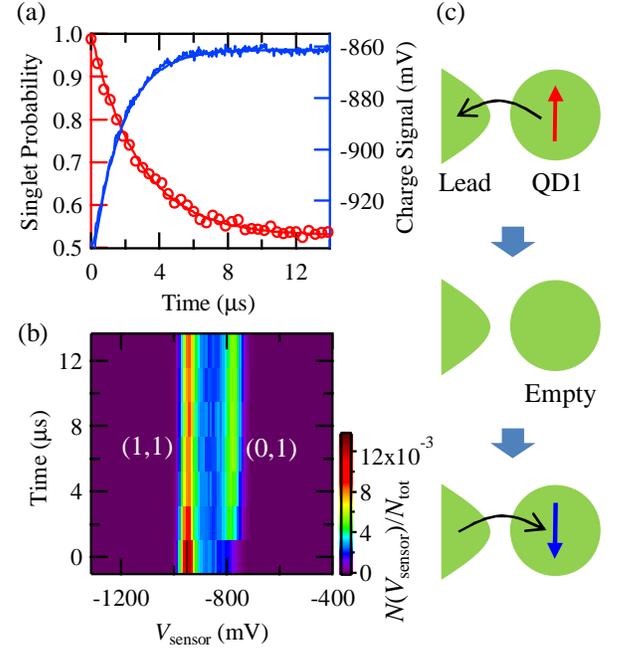}
  \caption{(a) Observed spin and charge signals (the singlet probability and the average of the sensor signal $\langle V_{\rm sensor} \rangle $) as a function of the interaction time.
  Red circles show the spin signal (left axis).
  The blue trace shows the charge signal (right axis).
  The smooth lines are exponential fits resulting in the relaxation time of 3.0~$\mu$s for the spin, and 1.8~$\mu$s for the charge.
  (b) Statistics of the charge signal at the operation point.
  Histogram of observed values of the charge sensor $V_{\rm sensor}$ (on the x axis), $N(V_{\rm sensor})/N_{\rm tot}$ is plotted as a function of the interaction time (y axis).
  The two peaks, at $V_{\rm sensor}=-960$ mV, and $-780$ mV, correspond to the (1,1) and the (0,1) charge states, respectively.
  The weight of the (0,1) component increases with the longer interaction time.
  (c) Schematic of the spin relaxation by a first-order tunneling process.
  An electron escapes from the QD and the QD becomes empty.
  Another electron comes in after that.
  }
  \label{CT}
\end{center}
\end{figure}

In this way, we first measure the spin relaxation using the operation point O$_1$ close to a charge transition line, see Fig.~\ref{Device}(b), where the QD level is close to the Fermi level of the lead.
The tunneling gate voltage is set to $V_{\rm T}=-660$~mV.
The red circles in Fig.~\ref{CT}(a) show the measured singlet probability as a function of the interaction time at O$_1$.
Initially at 1, the singlet probability decreases upon increasing the interaction time from zero.
This decrease indicates that a triplet component is formed by the interaction with the lead. Fitting with an exponential reveals a relaxation time of 3.0~$\mu$s.

Similarly to spin, we also measure the lifetime of charge in this configuration. 
The blue trace in Fig.~\ref{CT}(a) shows the averaged $ V_{\rm sensor} $ as a function of the interaction time. As seen there, 
$\langle V_{\rm sensor} \rangle $ changes exponentially, with the fitted charge relaxation time of 1.8~$\mu$s. To examine the charge relaxation in more detail, we plot in Fig.~\ref{CT}b histograms (the x axis) of the values of $V_{\rm sensor}$ for a varying interaction time (the y axis).
The two peaks along a horizontal cut correspond to the (1,1) and the (0,1) charge states, respectively.
At a zero interaction time, only the (1,1) state signal is present, while (0,1) state appears for finite interaction times.

In this configuration, the mechanism of the relaxation for both spin and charge is a first-order tunneling process~\cite{2015BiesingerPRL}. Namely, the electron tunnels out of the QD$_1$ into the lead, after which the dot is refilled from the lead, and the initial information is lost.
The spin and charge relaxation happen simultaneously, the information loss of the spin demonstrated in Fig.~\ref{CT}(a), and of the charge in Fig.~\ref{CT}(a-b). We note that though the relaxation timescales are similar, they are not identical. The difference comes from a difference in the rate dependence on Fermi occupation of the lead (see the Supplementary Information).


\begin{figure}
\begin{center}
  \includegraphics{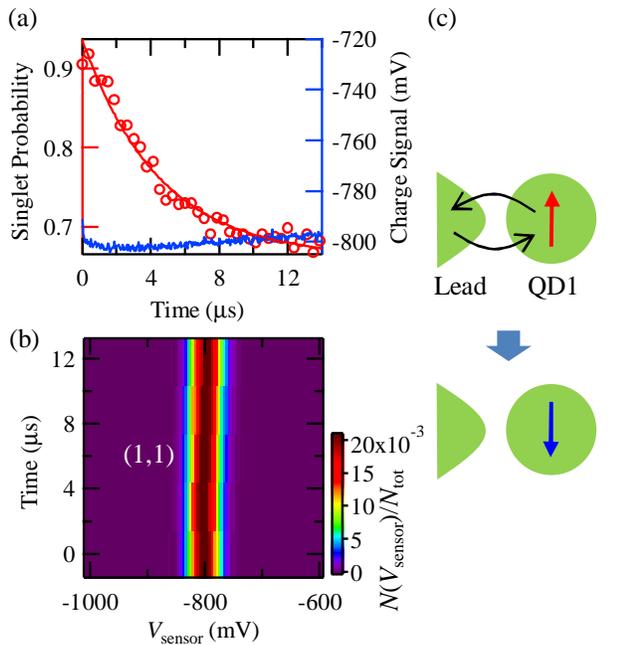}
  \caption{(a) The observed singlet probability and $\langle V_{\rm sensor} \rangle $ as a function of the interaction time at O$_2$ [see Fig.~\ref{Device}(b)].
  Red circles show the spin signal (left axis).
  The blue trace shows the charge signal (right axis).
  The red smooth curve is an exponential fit resulting in the relaxation time of 4.5~$\mu$s.
  The charge signal shows no relaxation.
  (b)   Histogram of observed values of the charge sensor $V_{\rm sensor}$ (on the x axis), $N(V_{\rm sensor})/N_{\rm tot}$, is plotted as a function of the interaction time (y axis).
  The peak corresponds to the (1,1) charge state.
  (c) Schematic of the spin relaxation by a second-order tunneling process.
  An electron of the QD$_1$ is swapped with a one in the lead in a single step.
  The spin state is changed even though the charge state is stable.
  }
  \label{CB}
\end{center}
\end{figure}

We now investigate the spin dynamics in a Coulomb blockaded dot. To this end, we repeat the previously described measurement using the operation point O$_2$, deep in the (1,1) region, see Fig.~\ref{Device}(b).
Here the QD level is far below the Fermi level of the lead.
To increase the speed of the lead induced spin dynamics on the dot, we increase the dot-lead tunnel coupling by setting $V_{\rm T}=-560$~mV.
As can be seen in Fig.~\ref{CB}(a), similarly as before, the spin state displays an exponential decay, with the relaxation time of 4.5~$\mu$s. However, now the charge signal barely changes, indicating that  
the charge state is not affected. (The slight change of the charge signal in Fig.~\ref{CB}(a) is caused by the distorted voltage pulses applied on P1 and P2.
Due to a cross-talk between the plunger gates and the sensor, the pulse distortion slightly affect the observed charge signal.)
This is confirmed by Fig.~\ref{CB}(b), where the histograms of the values of $V_{\rm sensor}$ display a single peak corresponding to the (1,1) charge state. 
The spin therefore decays at a fixed QD charge configuration.

\begin{figure}
\begin{center}
  \includegraphics{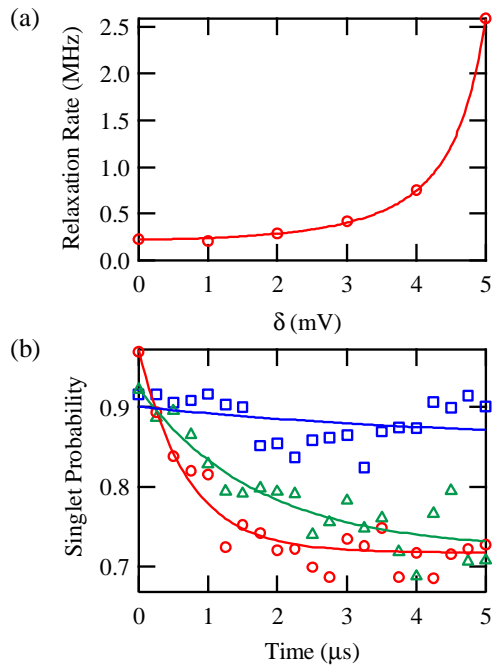}
  \caption{(a) The spin relaxation rate as a function of $\delta $.
  Circles show the experimental data and the line shows a theoretical curve considering the second-order tunneling process.
(b) The spin relaxation signal as a function of the interaction time at $O_2$ for different values of the gate voltage applied on gate T $V_{\rm T}$.
  Circles, triangles and squares show the result at  $V_{\rm T}=-560, -565, -570$~mV, respectively.
  The lines are exponential fits.
  }
  \label{Tune}
\end{center}
\end{figure}

We therefore interpret this as the observation of a spin relaxation induced by a second-order tunneling process~\cite{2005SchleserPRL}, where the electron in QD$_1$ swaps with a random one from the lead in a single step. 
Figure~\ref{Tune}(a) shows the spin relaxation rate as we change the operation point from O$_2$ toward O$_1$, parametrizing the displacement by voltage $\delta$. 
Upon increasing $\delta$ (moving towards the charge transition line) the spin relaxation rate is enhanced.
The measured dependence is very well fitted by an analytical expression for an inelastic cotunneling rate, giving $\propto (1/(\mu (2)-\mu _{F} )+1/(\mu _{F} -\mu (1)))^2$, with $\mu (N)$ and $\mu _{F}$being the electrochemical potential at the dot with $N$ electrons~\cite{2007HansonRMP} and the Fermi energy of the lead, respectively (see the Supplemental Information for details).
In addition to plunger voltages, we can tune the spin decay timescale by the voltage applied on gate T,  $V_{\rm T}$.
Indeed, as seen in Fig.~\ref{Tune}, applying more negative voltage $V_{\rm T}$ prolongs the spin relaxation time, by decreasing the tunnel coupling to the lead, from 0.7 to 1.7 to 5.0~$\mu$s, for $V_{\rm T}=-560, -565, -570$~mV, respectively.
(We note that the relaxation time at $V_{\rm T}=-560$~mV is different from the corresponding value of $V_{\rm T}$ given in Fig.~\ref{CB}(a) due to a shift of the QD conditions between experiments.)
This demonstrates the two handles on the speed of the lead induced dynamics of the QD spin.


To sum up the results observed in the Coulomb blockade regime, here the interaction with the lead influences only the dot spin, but not charge. 
The spin relaxation thus directly uncovers the second order tunneling processes.
This interaction can be utilized for the spin initialization, though also the measurement and manipulation might be envisioned, considering leads with special properties. 
We note that even though the timescale of the dot-lead interaction realized in this experiment was tuned to $\sim \mu$s, it is straightforward to enhance it by increasing the tunnel coupling, and/or utilizing the Kondo effect, which enhances the second-order tunneling at low temperatures.

In conclusion, we have measured spin dynamics in a QD-lead hybrid system.
Close to a charge transition, we observed spin and charge relaxation signals corresponding to the first-order tunneling process.
In the Coulomb blockade, we observed spin relaxation at a fixed charge configuration, corresponding to the second-order tunneling process.
The demonstrated dot-lead spin exchange can be useful as a general resource for spin manipulations, and simulations of open systems under non-equilibrium conditions.

\section{Methods}

The device was fabricated from a GaAs/AlGaAs heterostructure wafer with an electron sheet carrier density of 2.0~$\times$~10$^{15}$~m$^{-2}$ and a mobility of 110~m$^2$/Vs at 4.2~K, measured by Hall-effect in the van der Pauw geometry.
The two-dimensional electron gas is formed 90~nm under the wafer surface.
We patterned a mesa by wet-etching and formed Ti/Au Schottky surface gates by metal deposition, which appear white in Fig.~\ref{Device}(a).
All measurements were conducted in a dilution fridge cryostat at a temperature of 13~mK.

\section{Acknowledgements}

We thank J. Beil, J. Medford, F. Kuemmeth, C. M. Marcus, D. J. Reilly, K. Ono, RIKEN CEMS Emergent Matter Science Research Support Team and Microwave Research Group in Caltech for fruitful discussions and technical supports.
Part of this work is supported by the Grant-in-Aid for Scientific Research (No. 25800173, 26220710, 26709023, 26630151, 16H00817), 
CREST, JST, 
ImPACT Program of Council for Science, Technology and Innovation (Cabinet Office, Government of Japan), 
Strategic Information and Communications R\&D Promotion Programme, 
RIKEN Incentive Research Project, 
Yazaki Memorial Foundation for Science and Technology Research Grant, 
Japan Prize Foundation Research Grant, 
Advanced Technology Institute Research Grant, 
the Murata Science Foundation Research Grant, 
Izumi Science and Technology Foundation Research Grant, 
TEPCO Memorial Foundation Research Grant,
IARPA project ``Multi-Qubit Coherent Operations'' through Copenhagen University,  
DFG-TRR160, and the BMBF - Q.com-H  16KIS0109.

\section{Author contributions}
T. O., T. N., M. D., S. A., J. Y., K. T., G. A. and S. T. planned the project; T. O., T. N., M. D., S. A., A. L. and A. W. performed device fabrication; T. O., T. N., M. D., S. A., J. Y., K. T., G. A., P. S., A. N., T. I., D. L. and S. T. conducted experiments and data analysis; all authors discussed the results; T. O., T. N., M. D., S. A., J. Y., K. T., G. A., P. S. and S. T. wrote the manuscript.

\clearpage

\clearpage
\setcounter{equation}{0}
\setcounter{figure}{0}
\setcounter{table}{0}
\setcounter{page}{1}
\makeatletter
\renewcommand{\theequation}{S\arabic{equation}}
\renewcommand{\thefigure}{S\arabic{figure}}
\renewcommand{\bibnumfmt}[1]{[S#1]}

\onecolumngrid

\begin{center}
\textbf{Supplemental Material to `Tunneling induced spin dynamics in a quantum dot-lead hybrid system'}
\end{center}

\section{Charge and spin relaxation signals}

We describe the dynamics of the charge and spin on the QD by considering the rate equation
\begin{eqnarray}
\partial_t P_\sigma&=&-\Gamma_\sigma (1-f_\sigma)P_\sigma+\Gamma_\sigma f_\sigma P_{\rm e},
\label{eq:Ps}
\end{eqnarray}
for the probability $P_\sigma$ that the dot is occupied by a single electron with spin $\sigma \in \{ \uparrow, \downarrow\}$ (we alternative use $\sigma=\pm 1$), where $P_{\rm e}$ is the probability that the dot is empty. We do not consider any other states, which gives the normalization condition $P_\uparrow+P_\downarrow+P_{\rm e}=1$. Eq.~\eqref{eq:Ps} includes the process of an electron with a given spin leaving the dot into the lead where an empty state exists with the probability $(1-f_\sigma)$ and entering an empty dot from the lead state occupied with probability $f_\sigma$. This Fermi factor is given by $f_\sigma=f_{\rm FD}(\mu (1)-\sigma E_z)$,
with $\mu (1)=eV_g+\epsilon_1$ being the energy cost to add an electron into the dot, which includes the electrostatic potential energy $eV_g$, the orbital (quantization) energy $\epsilon_1$. The Zeeman energy is $E_z=g \mu_B B/2$, and the Fermi-Dirac distribution
\begin{equation}
f_{\rm FD}(\epsilon)=\left\{\exp\left[ \frac{\epsilon-\mu _F}{k_B T} \right]+1 \right\}^{-1},
\end{equation}
depends on the temperature $T$, and the lead Fermi energy $\mu _F$. Apart from the Fermi factors $f_\sigma$, the tunneling rates for hopping on and off the dot are identical, $\Gamma_\sigma$. We, however, allow for a spin dependence of the tunneling rate which has been found to be an appreciable effect (the asymmetry of the rates can be of the order of the rates themselves), most probably due to the exchange interaction in the lead~\cite{amasha2008:PRB,stano2010:PRB,yamagishi2014:PRB}.

To expose the spin and charge dynamics, we introduce new variables, the probability of charge occupation, $P_{\rm o}=P_\uparrow+P_\downarrow$ and the spin polarization, $s=P_\uparrow-P_\downarrow$, and new parameters, for the average, $\Gamma$, and the dimensionless asymmetry $\alpha$, in the tunneling rates, by writing $\Gamma_\sigma = \Gamma(1+\sigma \alpha)$, and similarly for the Fermi factors, $f_\sigma=f+\sigma f_\delta$. Equation \eqref{eq:Ps} can be now cast into the matrix form for the vector of unknowns, $v=(P_{\rm o},s)^{\rm T}$, namely
\begin{equation}
\partial_t v = - M (v-v_0),
\end{equation}
with the matrix defining the system propagator
\begin{equation}
M= 
\Gamma \left( \begin{tabular}{cc} 
$1+f+f_\delta \alpha$\, &\, $-f_\delta +(1-f)\alpha$\\
$f_\delta +(1+f)\alpha$\, &\, $1-f-f_\delta \alpha$\\
\end{tabular} \right),
\end{equation}
and the steady state solution
\begin{equation}
v_0=\frac{2}{1-f^2+f_\delta^2} 
\left( \begin{tabular}{c} 
$f(1-f)-f_\delta^2$\\
$f_\delta$
\end{tabular} \right).
\end{equation}
The steady state is independent on the tunneling rates, and depends only on the lead Fermi factors for the two spins, as it should be, while the propagator matrix depends on all parameters of the problem. Even though it is straightforward to solve the problem in the most general case, it is useful to consider $M$ for $\alpha=0$ (spin independent tunneling rates), which gives
\begin{equation}
M= 
\Gamma \left( \begin{tabular}{cc} 
$1+f$\, &\, $-f_\delta$\\
$f_\delta$\, &\, $1-f$\\
\end{tabular} \right).
\end{equation}
For a negligible difference of the Fermi function values for the two spins, $f_\delta\to0$, the charge and spin decay to their steady state values independently, with the rates $\Gamma(1+f)$, and $\Gamma(1-f)$, respectively. The steady states are also markedly different in this limit, as $P_{\rm o}(t=\infty) = 2f/(1+f)$ depends on the Fermi factors, while $s(t=\infty)=0$ does not. This is then the reason for difference in the decay scales: while the charge decays towards the steady state with effectively the sum of the rates for leaving, $(1-f)\Gamma$, and entering, $2f \Gamma$, the dot, only the events of electrons leaving the dots can relax the spin polarization $s$. The spin and charge relaxation scales will then be most different if $f\approx 1$, where the charge equilibrates much faster than the spin.

\begin{figure}[h]
\begin{center}
  \includegraphics{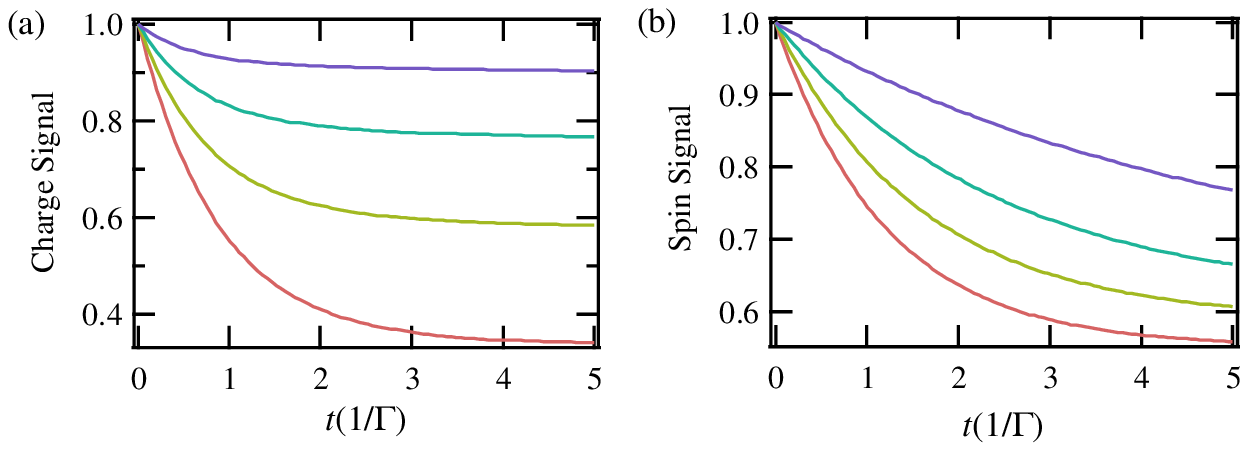}
  \caption{(a) Calculated charge signal as a function of $t$.
  The traces show the results with $f=0.2, 0.4, 0.6, 0.8$ from the bottom to the top.
  (b) Calculated spin signal as a function of $t$.
  }
  \label{relax}
\end{center}
\end{figure}

To demonstrate this difference, seen also experimentally, we plot the charge and spin signals in Fig.~\ref{relax} (a) and (b), respectively.
The parameters are set as $T=200$~mK, $B=0.5$~T, $g=-0.37$~\cite{2016OtsukaSciRep}.
The traces show the results with $f=0.2, 0.4, 0.6, 0.8$ from the bottom to the top.

\section{Co-tunneling rate}

To derive the formula for the spin relaxation by cotunneling, which was used in the main text to fit the data on Fig.~4(a), we consider the Hamiltonian of a QD coupled to a lead, $H=H_D+H_L+H_T$. Here the dot Hamiltonian is
\begin{equation}
H_D= \sum_{\alpha\in\{0,\sigma,S\}} \epsilon_\alpha |\alpha\rangle \langle \alpha |,
\end{equation}
where the index $\alpha$ labels the states of the dot $|\alpha\rangle$ with energies $\epsilon_\alpha$, and $|0\rangle$ denotes an empty dot, $|\sigma\rangle=d_\sigma^\dagger|0\rangle$ a dot with a single electron with spin $\sigma$, and $|S\rangle=d^\dagger_\uparrow d^\dagger_\downarrow|0\rangle$ a dot with a two electron singlet state, and $d^\dagger_\sigma$ is the creation operator of a dot electron with spin $\sigma$. The lead is described by
\begin{equation}
H_L=\sum_{k\sigma} \epsilon_{k\sigma} c^\dagger_{k\sigma} c_{k\sigma},
\end{equation}
where $k$ is a wave-vector (for simplicity, we consider a one dimensional lead, so that $k$ is a scalar). Finally, the dot-lead coupling is
\begin{equation}
H_T=\sum_{k\sigma} t_{k\sigma} c^\dagger_{k\sigma} d_\sigma + t^*_{k\sigma} d_\sigma^\dagger c_{k\sigma}, 
\end{equation}
which desctribes a spin-preserving lead-dot tunneling with, in general complex and spin and energy dependent, tunneling amplitudes $t_{k\sigma}$.

We now repeat the standard calculation \cite{averin1990:PRL, wegewijs2001:CM, golovach2004:PRB, lehmann2006:PRB, stano2015:PRB} with minor adjustments to to arrive at the inelastic spin decay rather than the co-tunneling current. To this end, we define the transition rate by the Fermi's Golden rule formula
\begin{equation}
\Gamma_{i\to f} = \sum_{i_{\rm lead}} p_{i_{\rm lead}} \sum_{f_{\rm lead}} \frac{2\pi}{\hbar} |\langle f_{\rm dot} \otimes f_{\rm lead} | G | i_{\rm dot} \otimes i_{\rm lead}\rangle|^2\delta(E_i-E_f),
\end{equation}
where $i$ and $f$ are the initial and final states with energies $E_i$ and $E_f$, respectively, considered to be separable (to the lead and dot components) eigenstates of the unperturbed system described by $H_0=H_D+H_L$. As we are not conditioning the transitions on the states of the lead, the rate is summed over all possible initial lead states, with the corresponding probabilities $p_{i_{\rm lead}}$, and all lead final states. The former gives the prescription for a replacement $\sum_{i_{\rm lead}} p_{i_{\rm lead}} |i_{\rm lead} \rangle \langle i_{\rm lead} | \to \rho^{\rm thermal}_{\rm lead}$, with the latter the equilibrium density matrix corresponding to a system with Hamiltonian $H_L$, at a temperature $T$. Finally, $G$ is the transition operator which can be expanded in powers of the tunneling term
\begin{equation}
G = H_T + H_T\frac{1}{E-H_0-i\gamma}H_T+ \ldots
\label{eq:G}
\end{equation}
with $E=E_i=E_f$. The two terms describe, respectively, the direct tunneling and the co-tunneling, and $\gamma$ is a regularization factor~\cite{begeman2010:PRB}.

Simple results can be derived in the well justified case of a negligible dependence of the tunneling amplitudes on the wave vector, $t_{k\sigma}\approx t_\sigma$. Using the first term in Eq.~\eqref{eq:G} gives in this limit the following expression for the direct tunneling rates defined in Eq.~\eqref{eq:Ps}
\begin{equation}
\Gamma_\sigma = \frac{2\pi}{\hbar} |t_\sigma|^2 g_F,
\end{equation}
where we denoted $\Gamma_\sigma \equiv \Gamma_{\sigma\to0} = \Gamma_{0\to\sigma}$ and $g_F$ is the density of states in the lead at the Fermi energy. Similarly, keeping only the second term in Eq.~\eqref{eq:G} gives the co-tunneling rate for a spin-flip (from $\sigma$ to the opposite value $\overline{\sigma}$) of a single electron occupying the dot,
\begin{equation}
\Gamma_{\sigma\to\overline{\sigma}} = \frac{\hbar}{2\pi} \Gamma_\sigma \Gamma_{\overline{\sigma}} 
\int {\rm d}\epsilon f_{\rm FD}(\epsilon+\sigma E_z) \left[1-f_{\rm FD}(\epsilon+\overline{\sigma} E_z)\right]
\left| \frac{1}{\epsilon-\mu (2)+i\gamma} -\frac{1}{\epsilon-\mu (1)-i\gamma}\right|^2,
\label{eq:aux}
\end{equation}
where $\mu (2)=\epsilon_S-\epsilon_1$ is the (spin independent part of the) energy cost to add a second electron into the dot. 
The expression can be further simplified if the dot is deep in the Coulomb blockade, so that the charge excitation energies are much larger than the temperature, namely $\mu (1) \ll \mu _F \ll \mu (2)$ are well fulfilled on the energy scale of the temperature, $k_B T$. The energy dependence of the last term in Eq.~\eqref{eq:aux} can be then neglected, replacing $\epsilon\to \mu _F$, and the remaining integral can be evaluated resulting in 
\begin{equation}
\Gamma_{\sigma\to\overline{\sigma}} = \frac{\hbar}{2\pi} \Gamma_\sigma \Gamma_{\overline{\sigma}} 
\frac{2\sigma E_z}{\exp\left(\frac{2\sigma E_z}{k_B T} -1\right)}
\left( \frac{1}{\mu (2)-\mu _F} +\frac{1}{\mu _F-\mu (1)}\right)^2,
\label{eq:final}
\end{equation}
where we also neglected the regularization factors. In the large temperature limit, $k_B T \gg E_z$, the temperature dependent factor becomes $k_B T$, while in the opposite limit, $k_B T \gg E_z$, it gives $2E_z$ for $\sigma=\downarrow$, and $0$ for $\sigma=\uparrow$. However, Eq.~\eqref{eq:final} is already in the form which was used to fit the data and is thus the final result of this section.

\end{document}